# Modeling of fracture geometry alteration and fracture flow evolution under geostress and water-rock interaction

Cheng Yu [1] (*)

*1. Key Laboratory of Hydraulic and Waterway Engineering of the Ministry of Education, Chongqing Jiaotong University, Chongqing, China*

(*)*: corresponding author, email: beimingyu@pku.edu.cn*

Abstract:

A coupled mech-hydro-chemical model for rock geometry alteration of fractures under water-rock interaction (WRI) and geostress is developed. Processes including WRI, asperity deformation, mineral chemical dissolution and pressure dissolution etc., are taken into account. A feature of this model lies in its approximate linearization to the non-linear pressure dissolution process, which makes this model compatible with existing numerical solute transport models. Case study shows that although usually only a thin layer of rock surface is invaded by WRI, the mechanical weakening of this thin layer tend to induce significant increase in rock surface deformation. Thus, the distributed flow field, mineral dissolution rates, and surface alteration increments etc., are all affected. This indicates that when fracture flow related issues are concerned, we should focus on the top thin layer of rock surface rather than the matrix. Results also show that fluid flow enhances the dissolution of rocks and development of flowing channels; however, at places where stress induces high pressure-dissolution but fluid flows slowly, minerals precipitate and fill up fracture voids. This suggests the hydraulic condition plays a key role in the development of fracture flow channels and the evolution of fracture geometry.

## 1. Introduction

The geometry alteration of rock fractures has been calling attentions from researchers of geosciences, and underground engineering etc. Because it determines the fracture morphology and related fracture flow and solute transport processes. The flow and transport processes have been discussed even more in various subjects such as $CO_2$ sequestration (e.g. Shukla et al. 2010), radioactive waste isolation (e.g. Rechard et al. 2013) and unconventional oil/gas exploration and production.

The mechanisms that alter the fracture geometry have been found to include mechanical deformation under geostress (e.g. Guo et al., 2014; Li et al., 2015), water-rock interaction (Karpyn et al., 2009; Zhang et al., 2015), mineral chemical dissolution (Detwiler & Rajaram, 2007; Ameli et al., 2014), and pressure dissolution (Bernabé & Evans, 2014; Lang et al., 2015). As the link between these processes, the fluid flow through rough fractures has been focused in many studies (Witherspoon et al., 1980; Zimmerman& Bodvarsson, 1996; Kohl et al., 1997; Sisavath et al., 2003; Qian et al., 2011; Yu 2015). The fluid flow mainly depends on the irregular fracture geometry, which is significantly affected by the rock surface dissolution and deformation. Conversely, the fracture geometry is also a result of fluid flow. Therefore, the fracture geometry alteration and the fracture fluid flow are highly coupled.

When fluid flows slowly but normal stress is high, pressure dissolution plays a more significant role. In geology, this process is usually employed to explain how stylolites are generated. However, when fluid flows fast, mineral dissolution at pores gets stronger, and thus fractures grow, just like the growth of karst caves.

Various numerical models have been developed to simulate the above mech-hydro-chemical process (e.g. Leem 1999; Watanabe 2005; Yasuhara, 2006; Detwiler & Rajaram, 2007; Taron et al., 2009; Ameli et al., 2014; Bernabé & Evans, 2014; Huang et al., 2015; Lang et al., 2015). However, their conceptual models are all more or less over-simplified. For example, some model simplified the mechanical model by employing uniform average stress at every contact grain, or simplified the chemical processes by only solely considering chemical dissolution. Moreover, existing models all regard the mechanical properties of rock surfaces homogeneous and insensitive to fluid, but virtually plenty of experiments show that WRI tends to cause asperity mechanical weakening (e.g. Hadizadeh & Law, 1991), which is an important impact factor of fracture deformation (e.g. Xie & Shao, 2006).

Therefore, the main purpose of this study is to develop a fuller functional model that takes all the above processes and impact factors into account. Moreover, fracture flow and mineral transport models are also improved, to provide better flow conditions for dissolution process. A numerical case study is carried out as a function check of our model. The impacts of WRI induced mechanical weakening, fracture roughness, and fluid flow velocity are examined. Results show that the fracture deformation is very sensitive to the mechanical property changes in the top thin layer of rock surface. Thus we should focus more on the top layer of asperities rather than the rock matrix when we are concerned with fracture flow related issues. It is also shown that when the pressure dissolution rate is high but the fluid flow is slow, minerals tend to precipitate from the fluid and fill up the fracture voids.

2. Methodologies

2.1 Rough surface generation

The rough fracture surface has been widely agreed fractal. The un-uniform fracture void space is mainly due to the mismatch of upper and lower surfaces of fractures, which are usually displaced by shear (Pyrak-Nolte and Morris, 2000; Schmittbuhl et al., 2008; Schwarz and Enzmann, 2012). Besides imaging technologies like CT scanning, numerical technics are often employed to generate fractal rough surfaces as well. In this study, we use the diamond-square method to generate the fracture aperture map.

2.2 Mechanical model considering WRI induced mechanical weakening

Many of the existing mechanical models for fracture surface deformation regard the rough surface as a composition of a semi-infinite extended rock matrix and a thin layer of non-uniformly distributed asperities. The semi-infinite rock matrix is usually considered elastic, but the thin layer of asperities is considered either rigid (Lang et al., 2015) or elastic (Pyrak-Nolte & Morris, 2000) in different studies.

In this study, the model by Pyrak-Nolte & Morris (2000) is followed. However, the asperity layer is further modified into a combination of weakened top-layer and unweakened sub-layer (figure 1). Like in other studies, asperities are assumed equivalent to a set of cylindrical columns, and in this case, the rock surface deformation includes both deformations of the semi-infinite matrix and the cylindrical columns. Mathematical solution of this model has been given by Pyrak-Nolte & Morris (2000), in which the major effort is to determine the rock matrix deformation under distributed normal stresses in terms of the Boussinesq's solution. The only difference of our model is the equivalent elasticities of cylinders.

A consideration about the asperity deformation has not been discussed yet in existing studies. It is about the selection of the boundary between semi-infinite matrix and the asperity layer (black dash line in figure 1). If supposing the normal stress constant, the position of the boundary does not affect the deformation of matrix, but does affect the deformation of single asperities. Therefore, the determination of the black dash line really matters, which was usually selected at the lowest of the rough surface (Petrovitch et al., 2014). In this case, the calculated total deformation is in fact an upper bond. On the other hand, if following Lang et al., (2015) to ignore the asperity deformation and only keep matrix deformation, the calculated total deformation turns out a lower bond. In this study, we follow the Pyrak-Nolte & Morris (2000), thus in fact our model tends to over-predict the deformation where normal stress is high.

The elastic moduli of cylinders should be modified into equivalent values of the double-layered asperity:

$$E_e = \left(\frac{h'}{h \cdot E'} - \frac{h' - h}{h \cdot E}\right)^{-1} \tag{1}$$

$$\nu_e = \left(\frac{\nu' h'}{E'} + \frac{\nu h}{E}\right) \cdot \frac{E_e}{h' + h} \tag{2}$$

where $E_e, \nu_e$ are equivalent Young's modulus and Poisson's ratio of the weakened asperity. $h$ is the thickness, $E$ is the Young's modulus, and $\nu$ is the Poisson's ratio of the unweakened sub layer, while $h', E', \nu'$ are those of the weakened top-layer.

Technically this mechanical model does not give an explicit solution of distributed stresses and deformations under certain total load. Instead, the program is de-signed to compress the fracture slowly down and monitor the total normal load. If the load exceeds the target, then restore the previous state and reduce the step size of the compression.

2.3 Fluid flow and mineral transport model

Fluid flow and mineral transport have been widely agreed as key driving forces to many geological processes. For long, has the fluid flow through rough fracture been studied and the "local cubic law" (LCL) has been used. It assumes the hydraulic head fits equation (3):

$$\nabla \cdot \left[ \frac{\rho_w g}{12\mu} b^2(\mathbf{x}) \cdot \nabla H \right] = 0 \tag{3}$$

where $\rho_w, \mu$ are fluid density and viscosity, and $b$ is local fracture aperture. However, its inaccuracy is still subject to debate. Wang et al., (2015) corrected the LCL for weak inertia and fracture tortuosity. However, we only follow their correction for fracture tortuosity, in consideration of better applicability.

In the modified LCL, local fracture transmissivity at cell (*i*,*j*) is evaluated in terms of equation (4):

$$\begin{cases} T_x^i = \dfrac{\rho g}{12\mu} \cdot \dfrac{2(b_x^{i+1})^3 (b_x^i)^3}{(b_x^{i+1})^3 + (b_x^i)^3} \\ T_y^j = \dfrac{\rho g}{12\mu} \cdot \dfrac{2(b_y^{j+1})^3 (b_y^j)^3}{(b_y^{j+1})^3 + (b_y^j)^3} \end{cases} \tag{4}$$

where $T_x^i, T_y^j$ are the transmissivities along *x* and *y* axes at cell (*i*,*j*),

$$\begin{cases} b_x^i = b \cdot |\sin \phi_x| \\ b_y^j = b \cdot |\sin \phi_y| \end{cases} \tag{5}$$

are corrected "direct aperture" in *x* and *y* orientations. The $\phi_x$ (or $\phi_y$) is the angle between *z* axis and the *x* (or *y*) component of local flow velocity. The $b_x^{i+1}, b_y^{j+1}$ are corrected apertures at adjacent cells. Then the flow field is governed by equation (6):

$$T_x |\sin \phi_x| \frac{\partial^2 H}{\partial x^2} + T_y |\sin \phi_y| \frac{\partial^2 H}{\partial y^2} = 0 \tag{6}$$

The ensuing solute transport is then a typical advection-dispersion process.

$$\frac{\partial c}{\partial t} = \nabla(D\nabla c) + \nabla(uc) + S + \Phi \tag{7}$$

where $c$ is the solute concentration, $D$ is the dispersion coefficient, $u$ is the local average flow velocity, $S$ is the rate at sources or sinks, and $\Phi$ is the kinetic reaction rate. The $D$ is approximately specified as the value for dispersion coefficient between parallel plates (Taylor, 1954; Aris, 1956):

$$D = D^* + \frac{u^2 b^2}{210 D^*} \tag{8}$$

where $D^*$ is the molecular diffusion coefficient of ions in fluid. Usually contact areas are regarded impermeable, but in fact, SEM still shows several microns' residual aperture at the contacts. The $S$ and $\Phi$ represent the mineral dissolution or precipitation.

2.4 Mineral chemical dissolution and precipitation

Chemical dissolution refers to the process that minerals are dissolved into fluid at free pores rather than at contacts. At free pores, the chemical dissolution rate *R* of minerals from pore surfaces into fluid is:

$$R = \frac{1}{b} \cdot k_{\text{eff}}(c_s - c) = \frac{2}{b_x + b_y} \cdot k_{\text{eff}}(c_s - c) \tag{9}$$

$$k_{\text{eff}} = \frac{k}{1 + \frac{2kb}{Sh D^*}} \tag{10}$$

where $k_{\text{eff}}$ is the effective reaction coefficient rate, $k$ is the mineral surface dissolution rate, $c$ is local mineral concentration, $c_s$ is the current mineral solubility, and *Sh*=4.86 is the Sherwood number.. The $k_{\text{eff}}$ depends not only on the chemical properties of minerals, but also the hydraulic conditions at interface (Detwiler & Rajaram, 2007). Typically when fluid flows fast, the $k_{\text{eff}}$ depends upon the mineral surface dissolution rate (reaction controlled); while when fluid flows slowly, $k_{\text{eff}}$ depends on the flow velocity (transport controlled).

Equation (10) gives a smooth transaction of $k_{\text{eff}}$ from reaction controlled condition ($k_{\text{eff}} = k$) to transport controlled condition ($k_{\text{eff}} = \frac{Sh D^*}{2b}$). Although experiments show that $k$ depends on many other conditions as well (Liu & Dreybrodt,

1997), we specify $k \approx 1 \times 10^{-4}$cm/s as a typical estimation following Detwiler & Rajaram (2007).

When $c > c_s$, the oversaturation results in mineral precipitation. The precipitation rate is considered the same as $k_{\text{eff}}$. More specifically, at free pores the $S$ and $\Phi$ in equation (7) are:

$$\begin{cases} S_1 = 0 \\ \Phi_1 = R \end{cases}, \quad \text{(at free pores)} \tag{11}$$

2.5 Pressure dissolution

Unlike at free pores, the pressure dissolution plays a more significant role at contacts. Physically, pressure dissolution is regarded due to chemical potential jump of minerals (Lehner & Bataille, 1984; Lehner, 1990; Lehner, 1995; Lehner & Leroy, 2004), and the instantaneous mass flux $J$ per unit contact area is:

$$J(\mathbf{x}) = \rho_s K_s \left[ \frac{\Omega_s}{k_B T} (\sigma(\mathbf{x}) - p_f) - \ln\left(\frac{c(\mathbf{x})}{c_s}\right) \right] \tag{12}$$

where $\sigma$ is the local stress, $\rho_s, K_s$, and $\Omega_s$ are the density, phenomenological rate coefficient of dissolution, and molecular volume of solid. The $k_B, T, c_0, p_f$ are the Boltzmann constant, temperature, equilibrium solute concentration, and the fluid pressure. If noting $c_{s,\sigma}$ as the solubility under $\sigma$ (Stumm & Morgan, 1996):

$$c_{s,\sigma} = c_s \exp\left(\frac{\sigma \Omega_s}{k_B T}\right) \tag{13}$$

if only the pressure dissolution is concerned, the $J$ is a source term in equation (7):

$$\frac{\partial c}{\partial t} = \nabla(D \nabla c) + \nabla(uc) + \frac{\rho_s K_s}{\rho_f b^*} \left[ \frac{\Omega_s}{k_B T} (\sigma - p_f) - \ln\left(\frac{c}{c_s}\right) \right] \tag{14}$$

where $b^*$ is the residual aperture at contacts.

Mathematically, the ln($c/c_s$) brings problematic nonlinearity. In fact, as $c_s < c < c_{s,\sigma}$, if $\bar{c} = c/c_s - 1$, then $\ln(c/c_s) = f(\bar{c}) = \ln(1 + \bar{c})$, $\bar{c} \in [0, \bar{c}^*]$, ($\bar{c}^* = \exp(\frac{\sigma \Omega_s}{k_B T}) - 1$). Then $f(\bar{c})$ can be linearized to $g(\bar{c})$ that goes through $(0, 0)$ and $(\bar{c}^*, c_{s,\sigma}/c_s)$:

$$\ln\left(\frac{c}{c_s}\right) \approx g(\bar{c}) = -\frac{1}{c_s} \frac{\sigma \Omega_s}{k_B T} \cdot \frac{1}{e^{\frac{\sigma \Omega_s}{k_B T}} - 1} \cdot (c_s - c) = -\alpha(c_s - c) \tag{15}$$

The difference $\delta = f - g$ gets maximum at $\bar{c}' = \frac{\bar{c}^*}{\ln(\bar{c}^* + 1)} - 1$. If $\Omega_s = 3.7 \times 10^{-29}$m$^3$ for wa-

ter-silica system (Lang et al., 2015) and $T$=298K, then when $\sigma$<90MPa, $\delta$<10%. If substituting $g$ for $f$ into equation (14), we have:

$$\frac{\partial c}{\partial t} = \nabla(D\nabla c) + \nabla(uc) + \frac{\rho_s K_s}{\rho_f b^*} \cdot \frac{\Omega_s}{k_B T}(\sigma - p_f) + \frac{\rho_s K_s}{\rho_f b^*}\alpha(c_s - c) \quad (16)$$

Therefore, under the above assumptions, the pressure dissolution at contacts is approximated to equation (7) with respect to the following arguments:

$$\begin{cases} \Phi_2 = \frac{\rho_s K_s}{\rho_f b^*} \cdot \alpha(c_s - c) \\ S_2 = \frac{\rho_s K_s}{\rho_f b^*} \frac{\Omega_s}{k_B T}(\sigma - p_f) \end{cases} \text{(at contacts)} \quad (17)$$

Note that the linearization of ln($c/c_s$) ensures $c \leqslant c_{s,\sigma}$, but slightly underestimates the reaction term $\Phi_2$. In this case, equation (16) virtually predicts slightly higher concentrations at the contacts.

As the net mass flux from the rock surface into fluid is $J_{\text{net}}$, the dissolution induced alteration rate of asperity length $h$ is estimated:

$$\frac{dh}{dt}(\mathbf{x}) = \frac{J_{\text{net}}(\mathbf{x})}{\rho_s} \quad (18)$$

3. Case study

In this study, we construct and couple three sub models: 1) the mechanical model determines the distributed stress and deformation of surface; 2) the fluid flow model gives the flow field; and 3) the mineral transport model predicts the distributed mineral concentration and dissolution rates. These models are coupled by updating fracture asperities and apertures (figure 2). As the mechanical model is constructed in lattice grids, for compatibility, finite difference method based programs *MODFLOW* and *MT3DMS* are selected to do the modeling.

A case study is shown in figure 3, the grid of which has 50 rows and 50 columns with width 2 mm. Fluid flows right to left. Left and right boundaries have constant heads with drawdown 0.5mm, while the others are no-flow. The residual aperture at contacts is $b^*$=0.1μm. In practice, each transport step is modeled in "steady state mode" of *MT3DMS*, for the long-term semi-steady mineral concentrations under each geostress condition.

Model parameters are almost following those in previous studies, or estimated according to typical geological conditions: $E$=20GPa, $v$=0.25, $c_s$=5.9×10$^{-9}$%, $\rho_s$=2.71g/cm$^3$, $\Omega_s$=3.7×10$^{-20}$mm$^3$, $Sh$=4.86 (Detwiler & Rajaram, 2007), $K_s$=3.4×10$^{-12}$mm/h (Lang et al., 2015), $k$=3.6mm/h (Liu & Dreybrodt, 1997), for CaCO$_3$.

4. Results and analysis

As we do not have any experimental result, this case study is only a function check of our model. Results under different conditions are compared to examine the surface alteration and its impact factors.

4.1 Impact of WRI induced mechanical weakening

To validate the impact of mechanical weakening, two scenarios are compared. Apertures, mechanical properties, and hydraulic conditions all follow the above statement. However, one of the scenarios is not weakened ($E'=E$), while the other is ($E'=0.1E$). The normal load on the surfaces is assumed 150kN. Then the stress distribution, hydraulic conductivity, and surface alterations are examined.

Figure 4-6 show the initial status (first loop of figure 2) of apertures, semi-steady mineral concentrations, and surface dissolution rates of both scenarios.

Figure 4 shows that although only a very thin layer of rock surface is weakened, apertures turn out significantly affected. In weakened scenario, apertures are smaller and the contact area nearly doubles. This suggests that for fracture flow related issues, more attention should be paid on the top thin layer of rock surface, rather than the rock matrix, which might be very different. Additionally, fluid flows much slower when rock surface is weakened. As a result, mineral dissolution rate is much smaller as well.

Figure 6 shows both chemical dissolution rate at free pores and pressure dissolution rate at contacts. Three details can be recognized. 1) In un-weakened scenario, pressure dissolution and chemical dissolution nearly play equal roles in surface alteration, while in weakened scenario pressure dissolution is much smaller. 2) In the un-weakened scenario, larger area of rock surface is dissolving with higher rate, be-cause of the faster fluid flow. 3) Mineral precipitation in weakened scenario is heavier than that in un-weakened scenario. Dissolution rate is higher at the inlet because of larger under-saturation. On the contrary, precipitation always occurs near the contact areas where mineral concentration is high but fluid velocity is slow.

Figure 7 shows the fracture geometries of two typical sections (see figure 2). Section *A-A'* goes through the area with both significant chemical dissolution and pressure dissolution. In figure 7.a, at some locations newer profile is lower than older ones, because chemical dissolution at these places is smaller than nearby pressure dissolution. Fracture geometry alteration is clearly identified, especially near the inlet. Section *B-B'* represents the area where minerals precipitate. In figure 6.b, precipitation induced rock surface increment is identified in the middle where pressure dissolution is high but flow velocity is slow. Compared with the un-weakened scenario, geometry alteration in the weakened scenario is much smaller.

Figure 8 shows the flow rates under both scenarios. For both scenarios, the flow rates increase significantly, because chemical dissolution helps in the development of flowing channels. It is shown that the surface mechanical weakening significantly affects the alteration of rock surface. Although only a very thin layer of rock surface is invaded by the water, apertures and fluid flow field are strongly influenced. It indicates when we are concerned with the issues related to fracture flow, we should pay special attention to the top thin layer of the rock surface, rather than rock matrix, which virtually is much less sensitive to the fluid.

4.2 Impact of surface roughness

The impact of rock surface roughness is investigated as well. As a comparison, fracture asperities of the existing fracture model are doubled/halved to generate fracture geometries that have larger/smaller roughness. Since other conditions keep the same with the unweakened scenario in last section of this paper, model results to some extent represent the impact of rock surface roughness.

Figure 9 is the dissolution rates of rock surfaces with different roughnesses. With larger roughness, normal load is supported by fewer asperities, and most flow channels are not affected. Conversely, with smaller roughness, flow channels shrink, and result in smaller dissolution rates.

4.3 Impact of fluid flow velocity

Fluid flow plays a key role in the dissolution and precipitation of minerals. The competition between fluid transport capacity and the over-saturation of minerals determines whether flow channels are dredged up or filled up. In this section, the fluid velocity is adjusted by changing the hydraulic drawdown between inlet and outlet. Two numerical tests are carried out using hydraulic drawdowns 0.1mm and 1mm respectively, and the semi-steady concentrations, dissolution rates, and aperture alterations are examined.

Figure 10 is the variation of dissolution rate for flow fields with 0.1mm and 1mm hydraulic drawdowns. Dissolution rates at three moments at represented. When fluid flows fast (figure 10.a-c), generally dissolution rates increase continuously because of higher transport capacity of the fluid. But dissolution rates at the contact locations decrease, because pressure dissolution leads to larger contact areas. Note that in figure 10.a all dissolution rates above 0.085mm/h are still painted in red for better differentiation of other values. As the increase of fluid transport capacity, precipitation areas shrink continuously. Figures 10.a to 10.c imply that when fluid flows fast, flowing channels develop, and fracture gets higher conductivity.

On the contrary, the story for the other scenario is different. As the fluid transport capacity is low, at most areas of the rock surface, dissolution rate is very small. Precipitation occurs at large areas near the contacting locations, and these areas expand continuously. In this case, at these places fracture void is being filled up, and apertures decrease. The rock surface alteration can be recognized from the aperture profiles as well (figure 11).

Figure 11 shows the evolution of aperture profiles along the two typical sections, in which the solid lines go upward, while the dash lines move downward. The variation of solid lines is obvious, but little difference can be recognized between dash lines, because of the very slow fluid velocity and transport capacity.

The different directions of fracture development of the two scenarios can be seen from their mean transmissivities as well, which are estimated in terms their total flux $Q$, and their hydraulic drawdown $\Delta H$,

$$T = \frac{Q}{\Delta H} \tag{19}$$

Figure 11 gives the changes in fracture transmissivity. The increase of fracture transmissivity under fast flowing condition indicates the expansion of flow channels, while the transmissivity decrease when fluid flows slowly indicates the shrinkage of flow channels.

5. Conclusions and Discussions

This study shows a modified approach for modeling of fracture geometry alteration under various mechanical and hydraulic conditions. This approach accounts for a mech-hydro-chemically coupled processes. Factors including the WRI induced mechanical weakening, mechanical deformation of rock surface, chemical dissolution of minerals at free pores, and the pressure dissolution of asperities at contacts are all taken into account. A feature of this modified approach lies in the simplification and approximation of the mathematical model for pressure dissolution, such that the modified model can then be solved in current modeling tools like MT3DMS.

A case study shows that although usually only a very thin layer of rock surface is affected by WRI, the mechanical weakening of this thin layer probably causes significant impact on the fracture deformation. As a result, the contact areas, flow channels, distributed stresses, flow field, and dissolution rates etc. are all affected. This indicates that when we are concerned with the fracture flow, we should focus more on the top thin layer of fracture surfaces, which might be very different from that of the matrix.

Results also show that the key factor that almost determines the development of fracture is the hydraulic condition, or more specifically the balance between the fluid transport capacity and the pressure dissolution. Where fluid flows fast, minerals are transported away as soon as they dissolve, however, at places where stress induces high pressure dissolution and fluid flows slowly, minerals are likely to precipitate and fill the fracture void.

Acknowledgement:

This study is financially supported by National Science Foundation of China (No. 51409028).

Figures:

Figure 1. Modification to the mechanical model.

Figure 2. Flow chart of this study.

Figure 3. Original aperture map of the modeled rough fracture. Sections A-A' and B-B' are selected to compare how rock surface profile alters.

Figure 4. Aperture map of fracture under normal load.

Figure 5. "Steady state" mineral concentration map of minerals in fracture flow.

Figure 6. Dissolution rate map of rock surface under normal load.

Figure 7. Aperture alteration for both weakened and unweakened scenarios.

Figure 8. Transmissivity increases because of dissolution induced flow channel expansion.

Figure 9. Dissolution rate map for fracture surface with different roughness.

Figure 10. Dissolution rate alteration for flow fields with 0.1mm and 1mm hydraulic drawdowns.

Figure 11. Aperture alterations when hydraulic drawdown equals 0.1mm and 1mm. (precipitation area expands continuously in dash lines)

Figure 12. Fracture transmissivity alteration under hydraulic drawdown 1mm and 0.1mm.

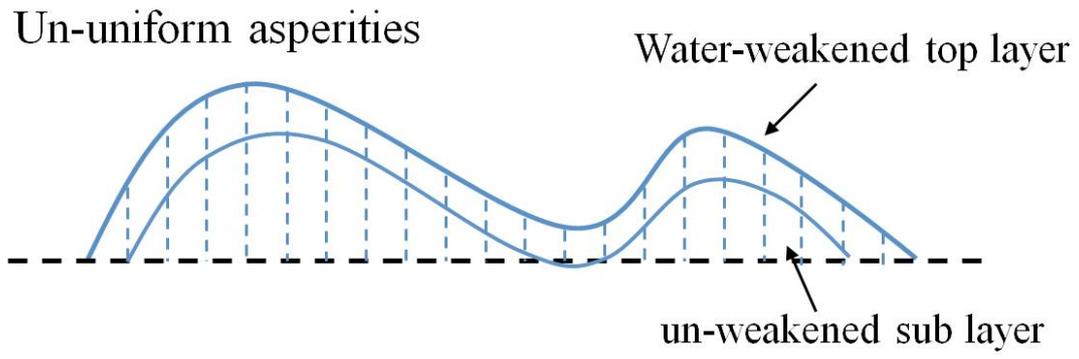

Figure 1. Modification to the mechanical model.

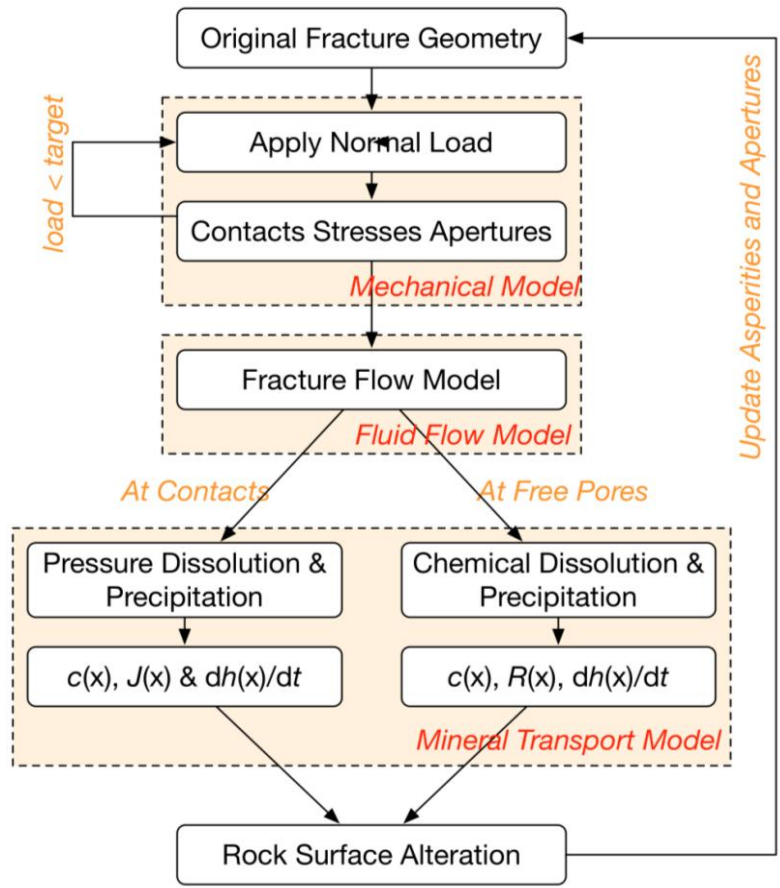

Figure 2. Flow chart of this study.

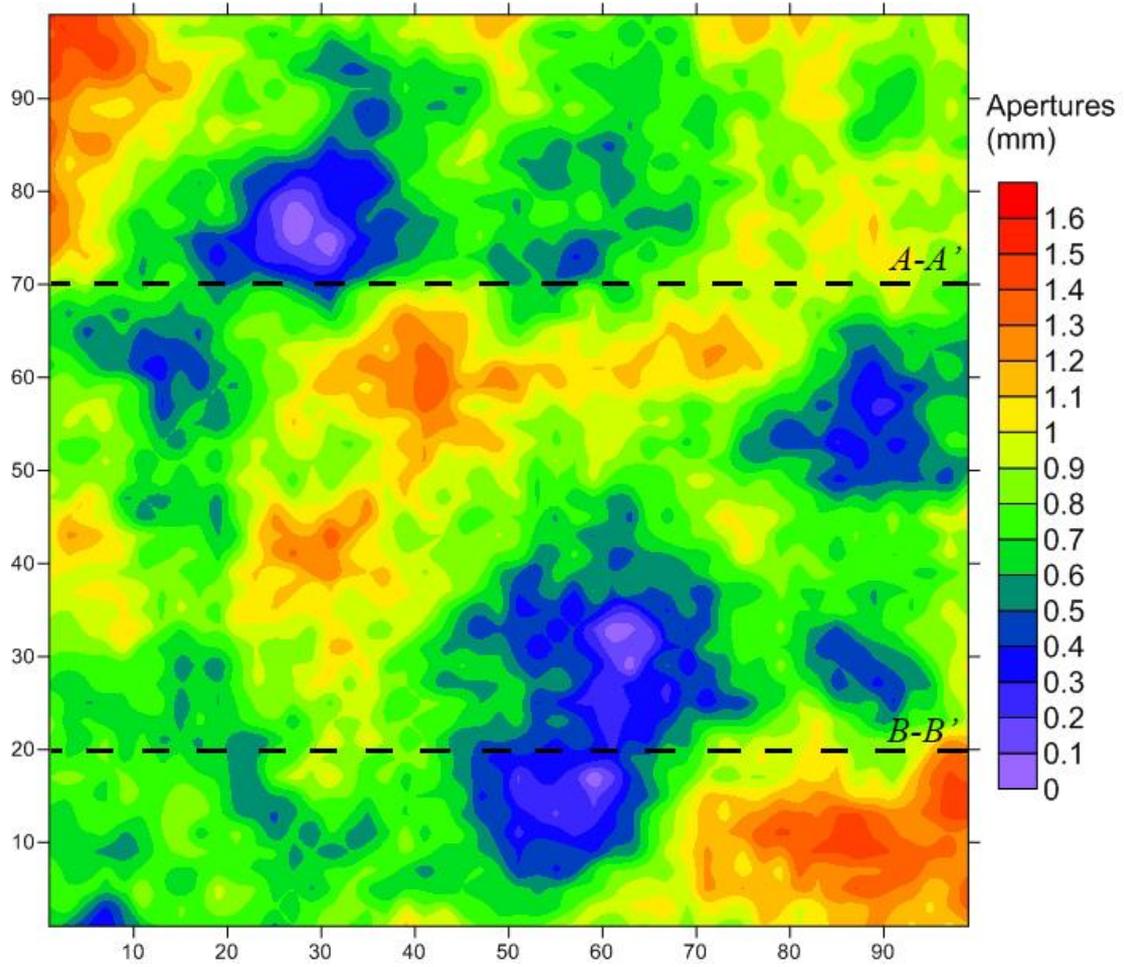

Figure 3. Original aperture map of the modeled rough fracture. Sections *A-A'* and *B-B'* are selected to compare how rock surface profile alters.

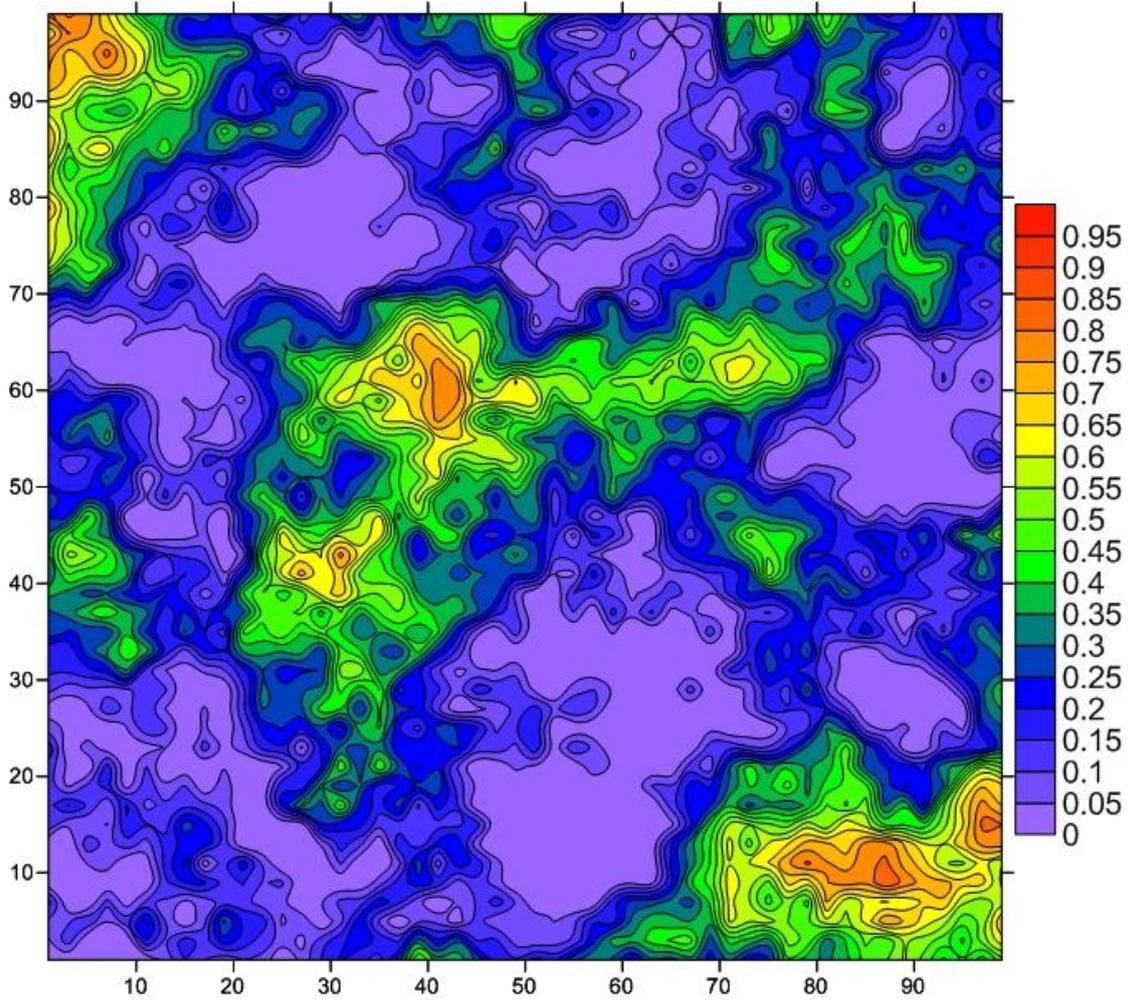

a. for weakened scenario

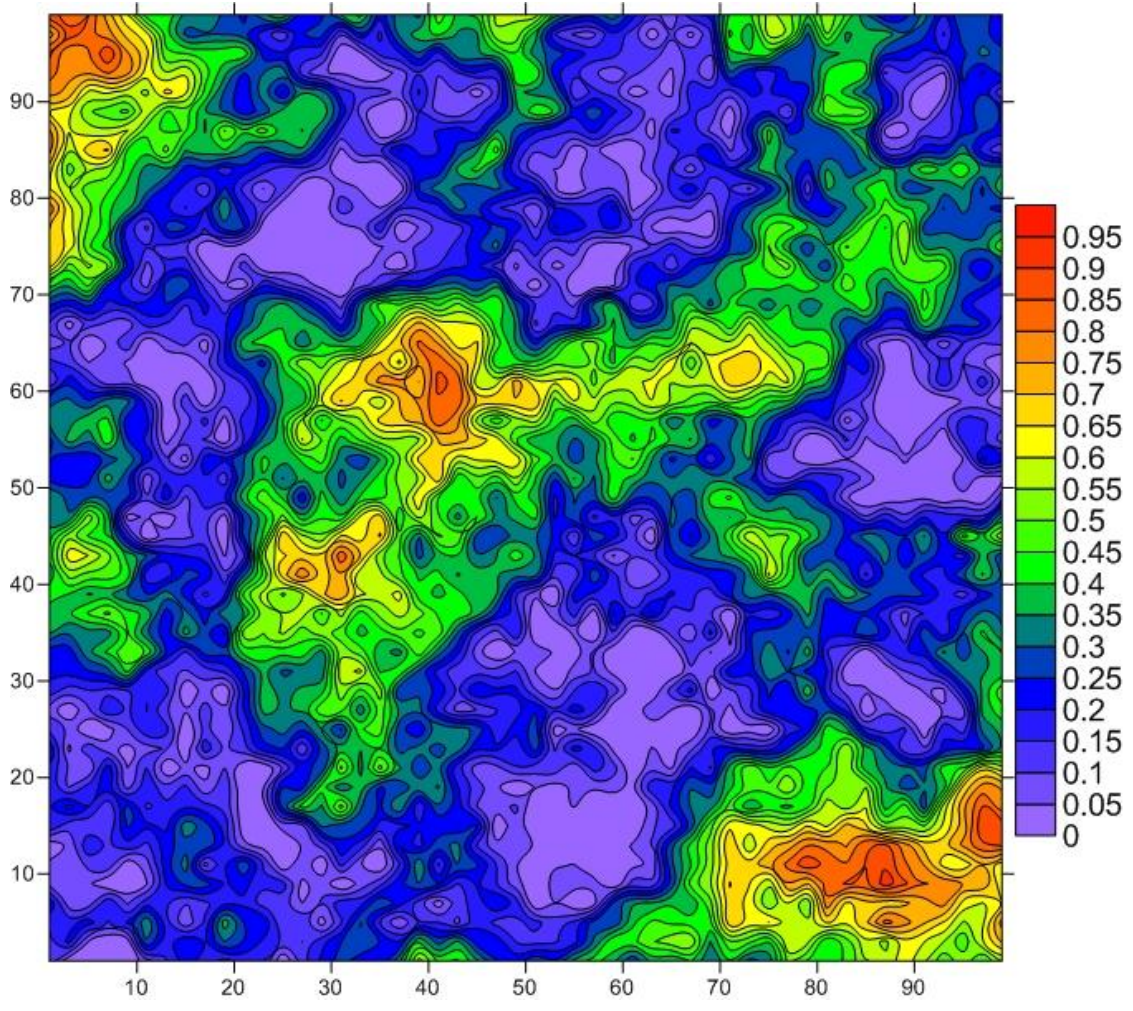

b. for unweakened scenario

Figure 4. Aperture map of fracture under normal load.

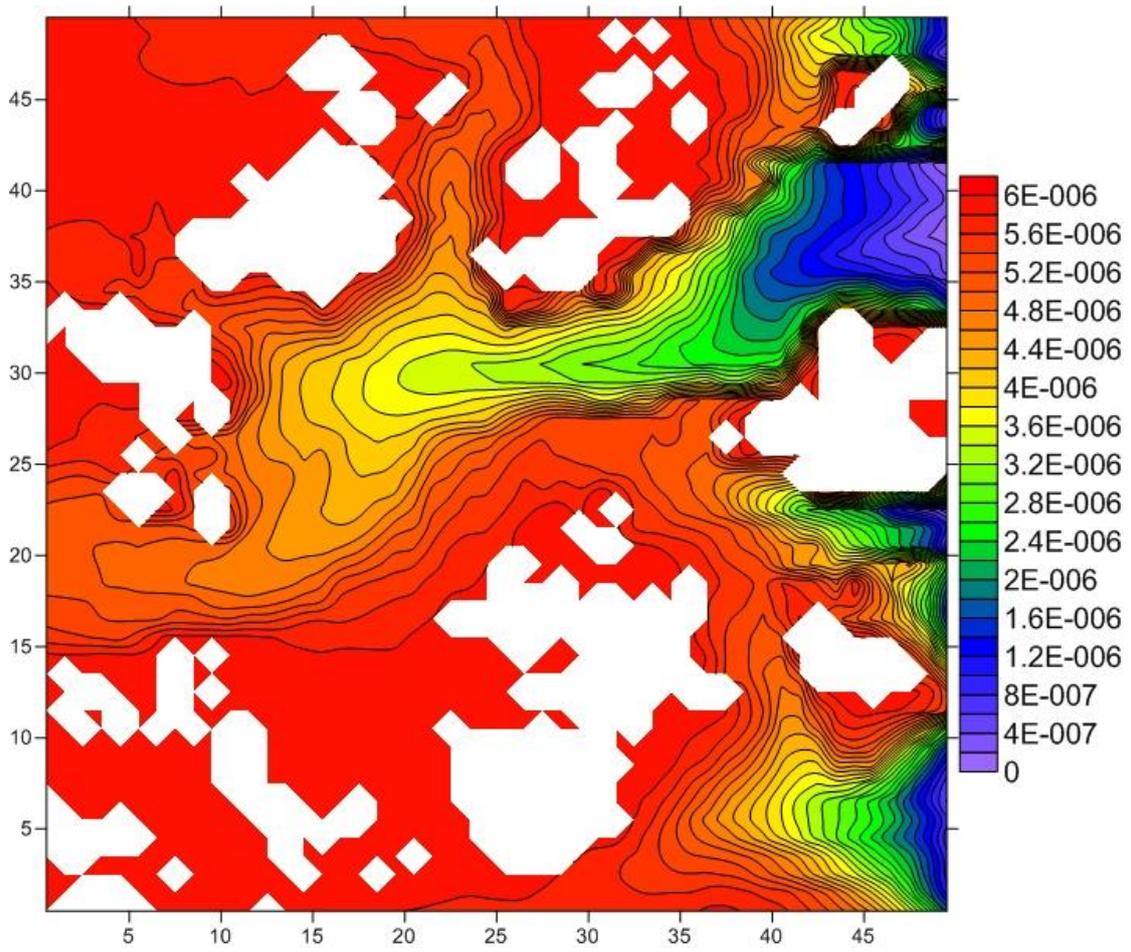

a. for weakened scenario

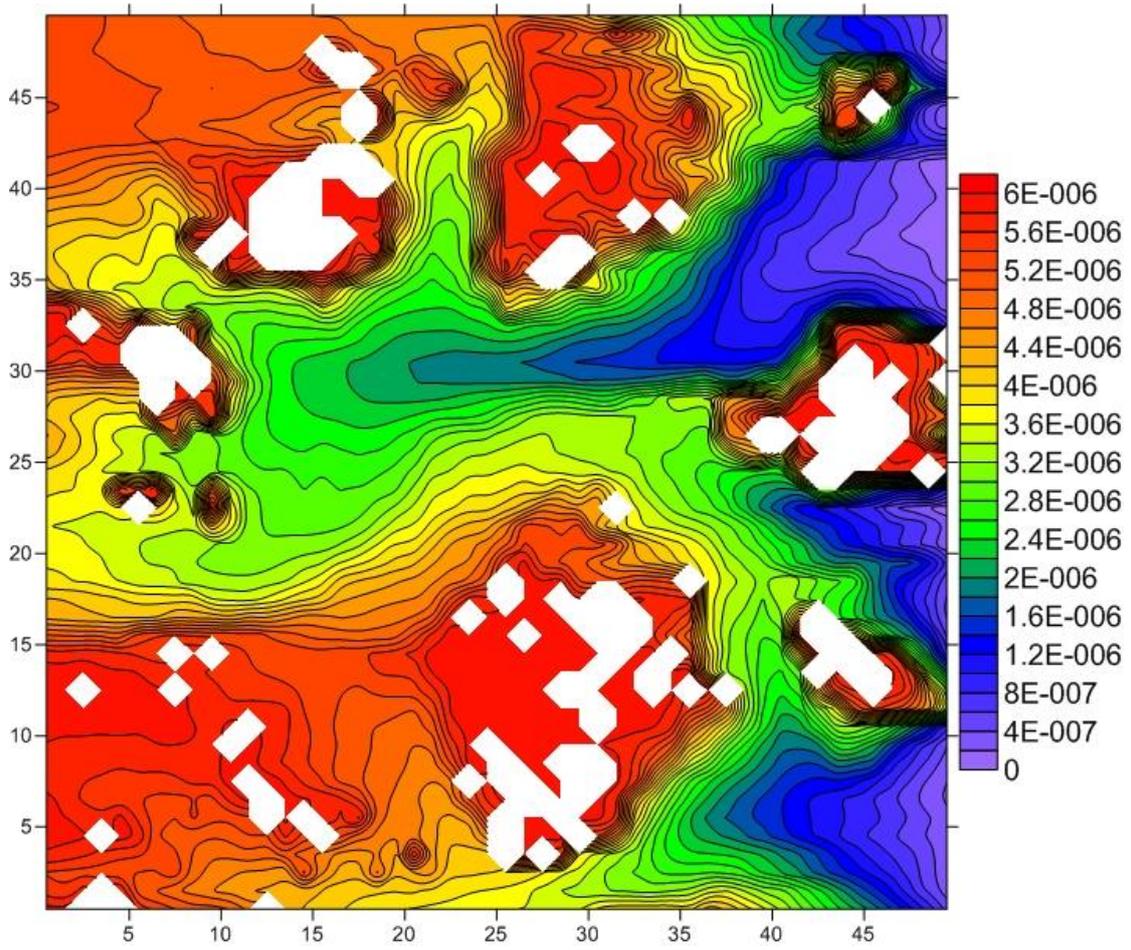

b. for unweakened scenario

Figure 5. "Steady state" mineral concentration map of minerals in fracture flow.

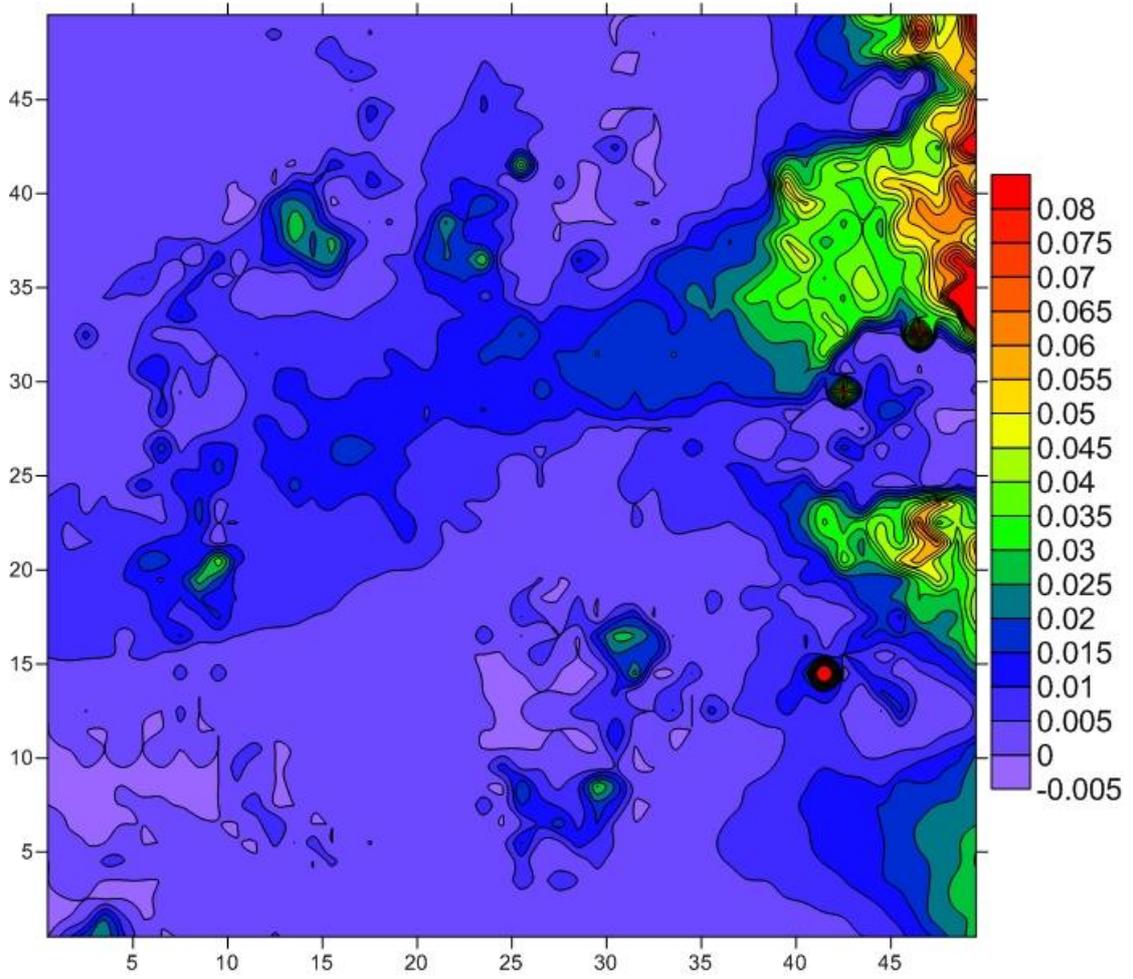

a. for weakened scenario

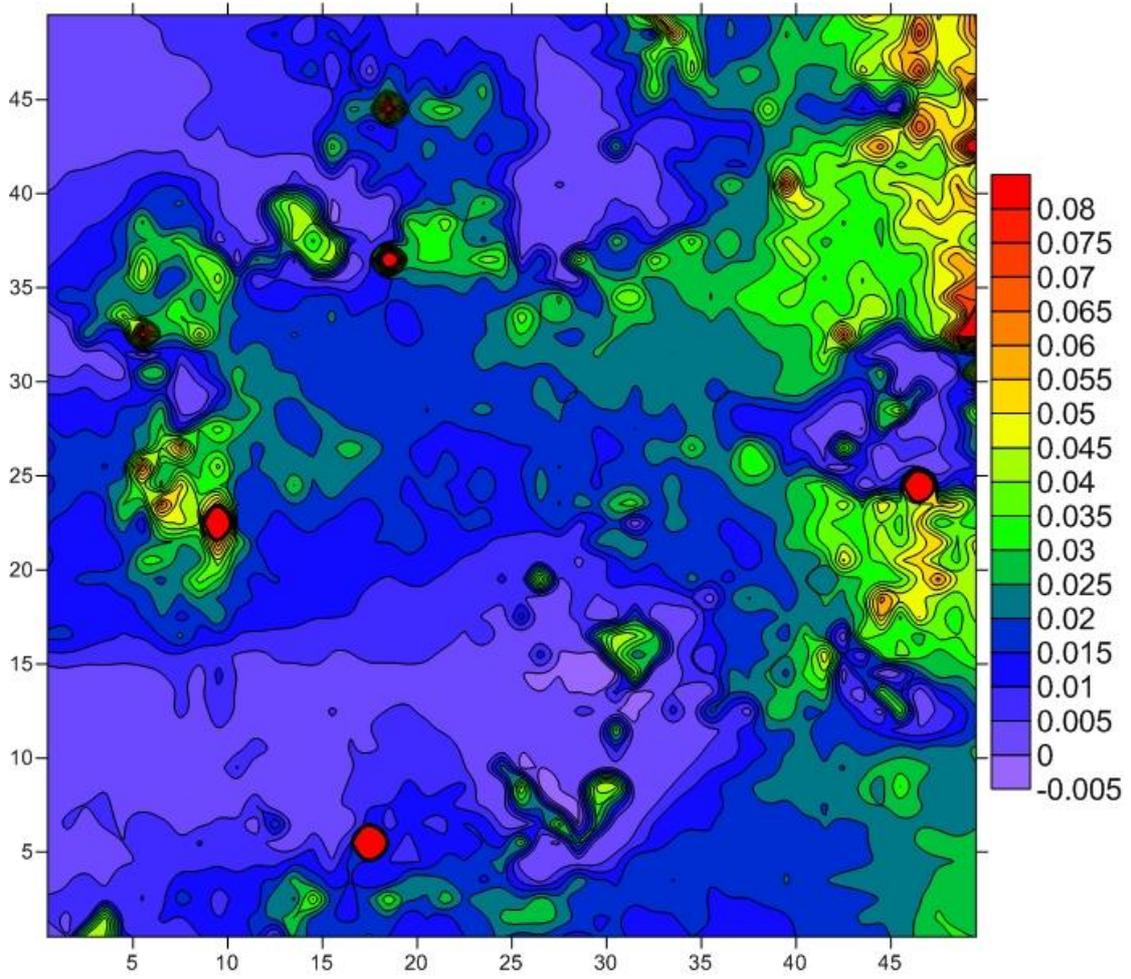

b. for unweakened scenario

Figure 6. Dissolution rate map of rock surface under normal load.

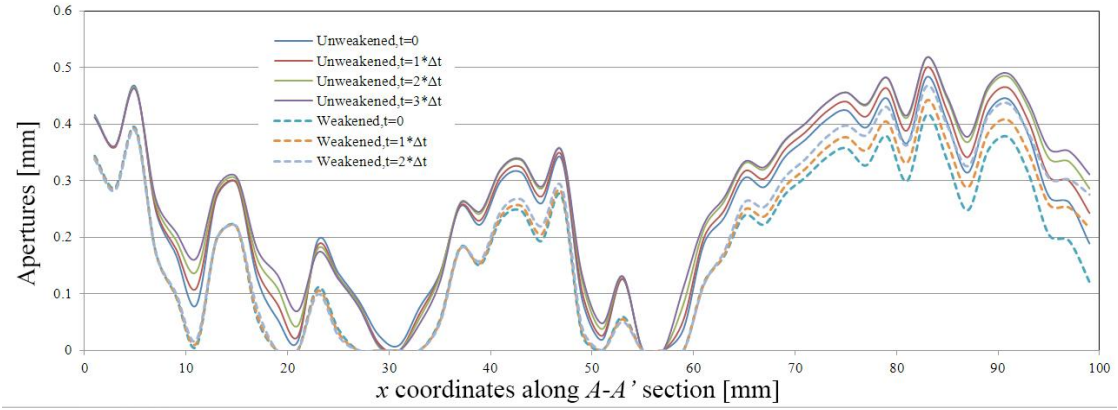

a. along *A-A'* section

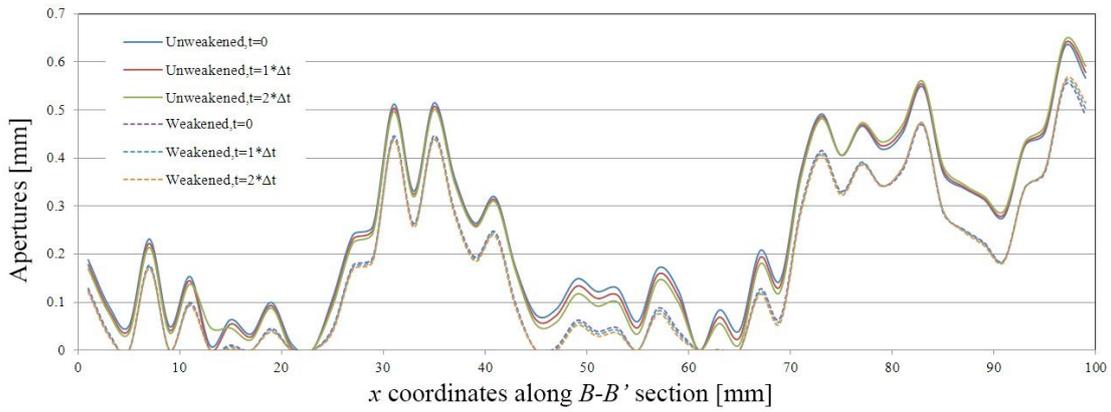

b. along *B-B'* section

Figure 7. Aperture alteration for both weakened and unweakened scenarios.

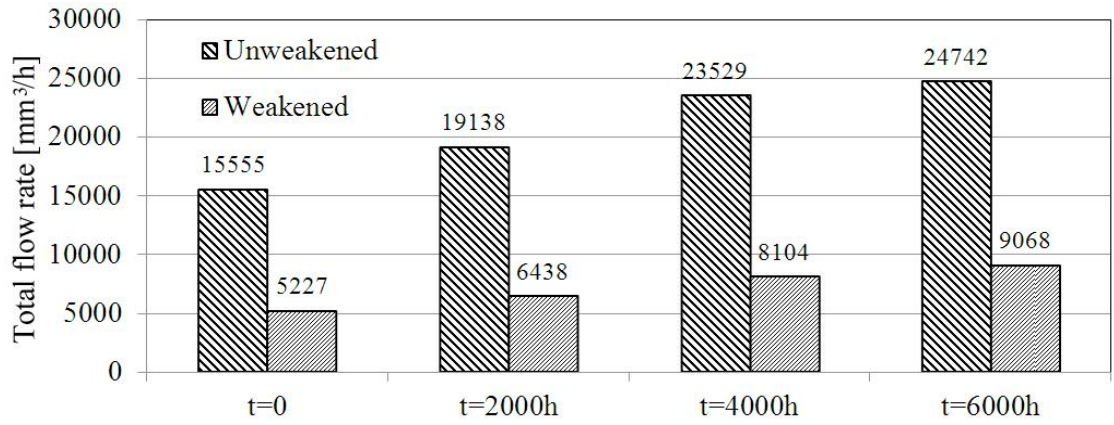

Figure 8. Transmissivity increases because of dissolution induced flow channel expansion.

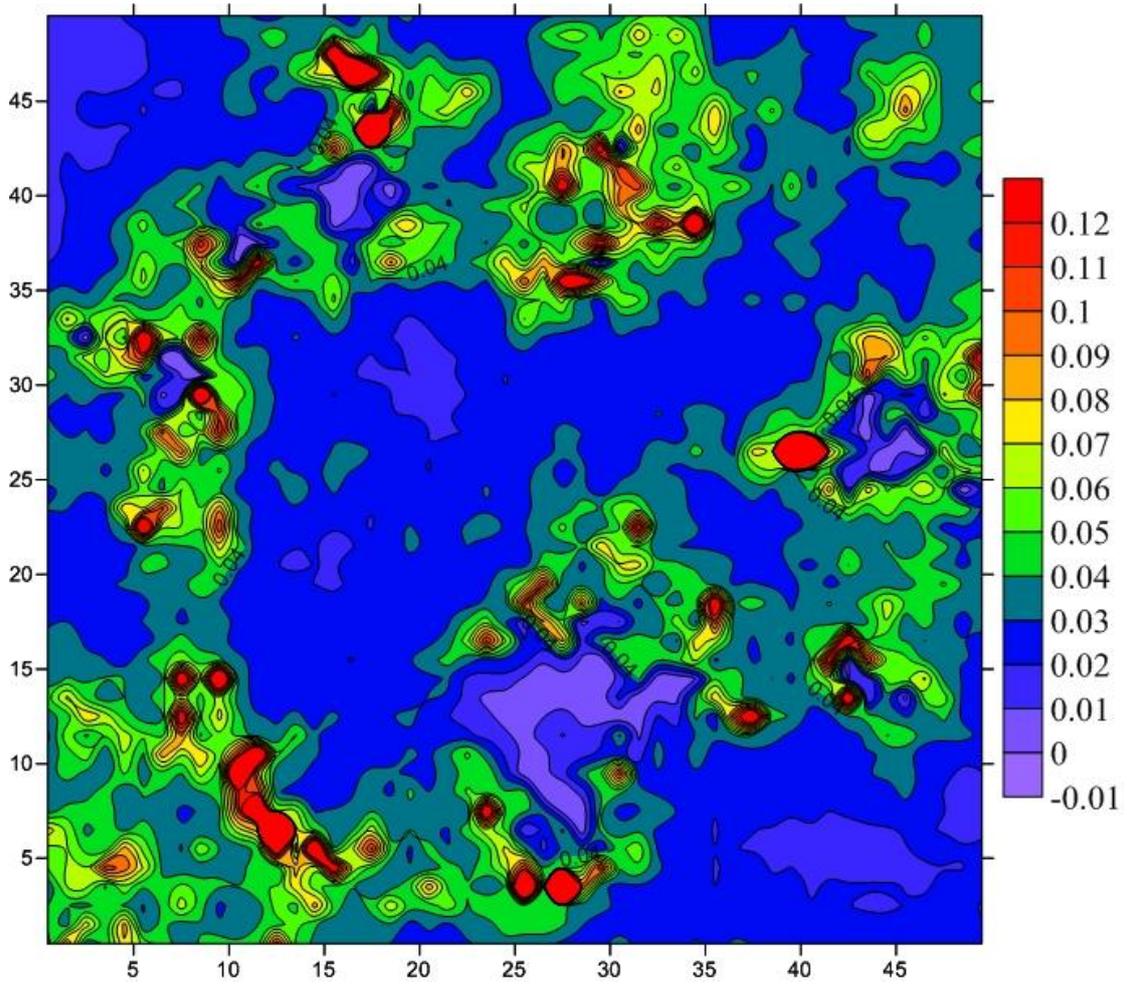

a. for fracture with double roughness of the original geometry

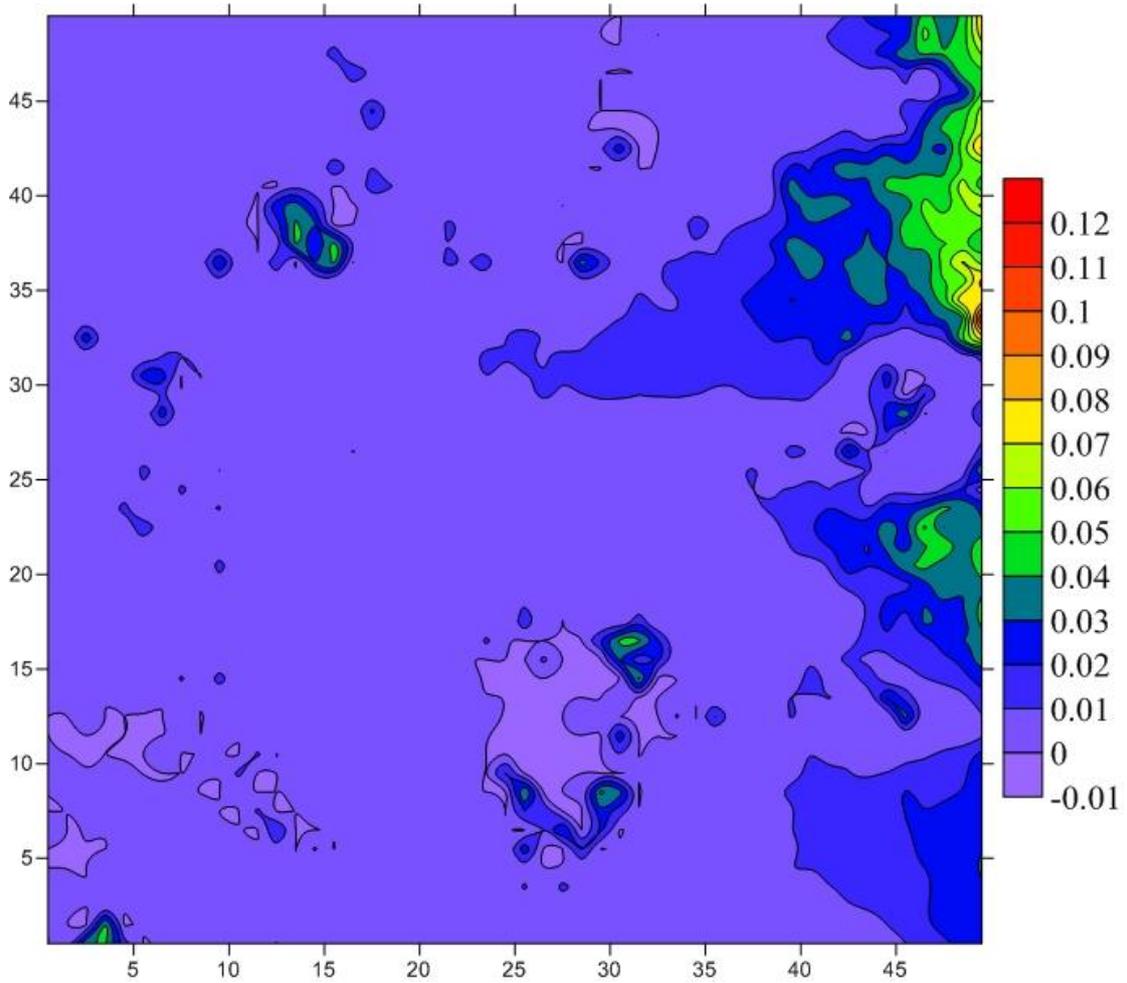

b. for fracture with half roughness of the original geometry

Figure 9. Dissolution rate map for fracture surface with different roughness.

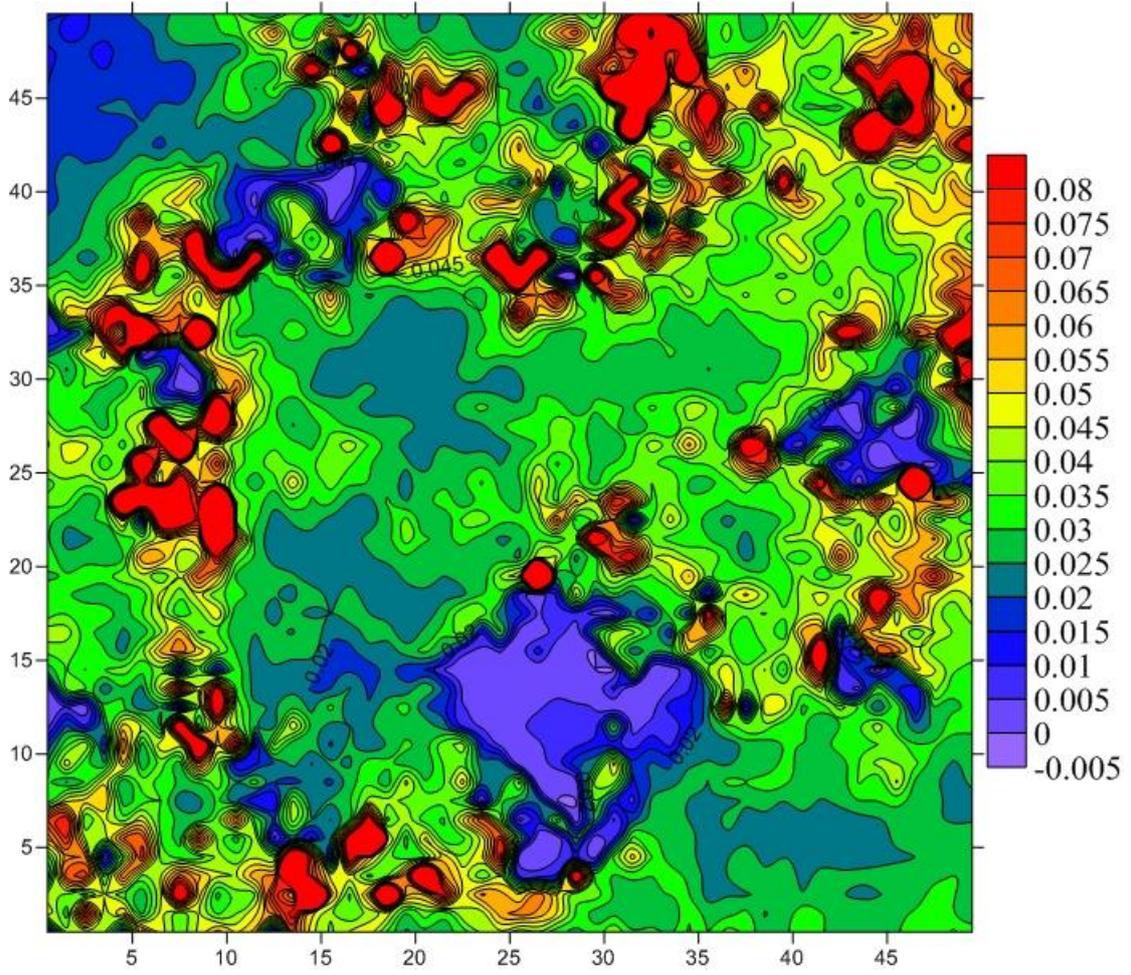

a. ΔH=1mm, t=0

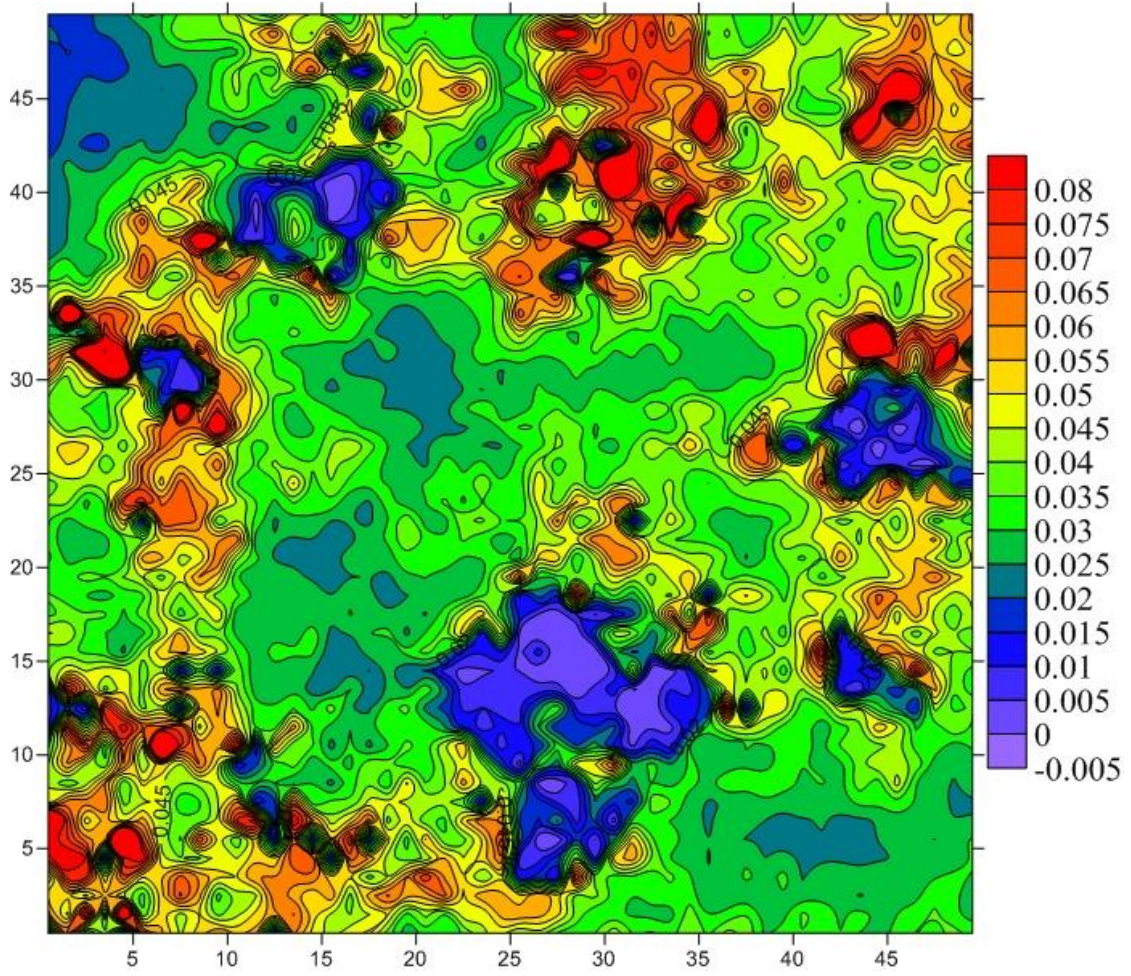

b. Δ*H*=1mm, *t*=2000h

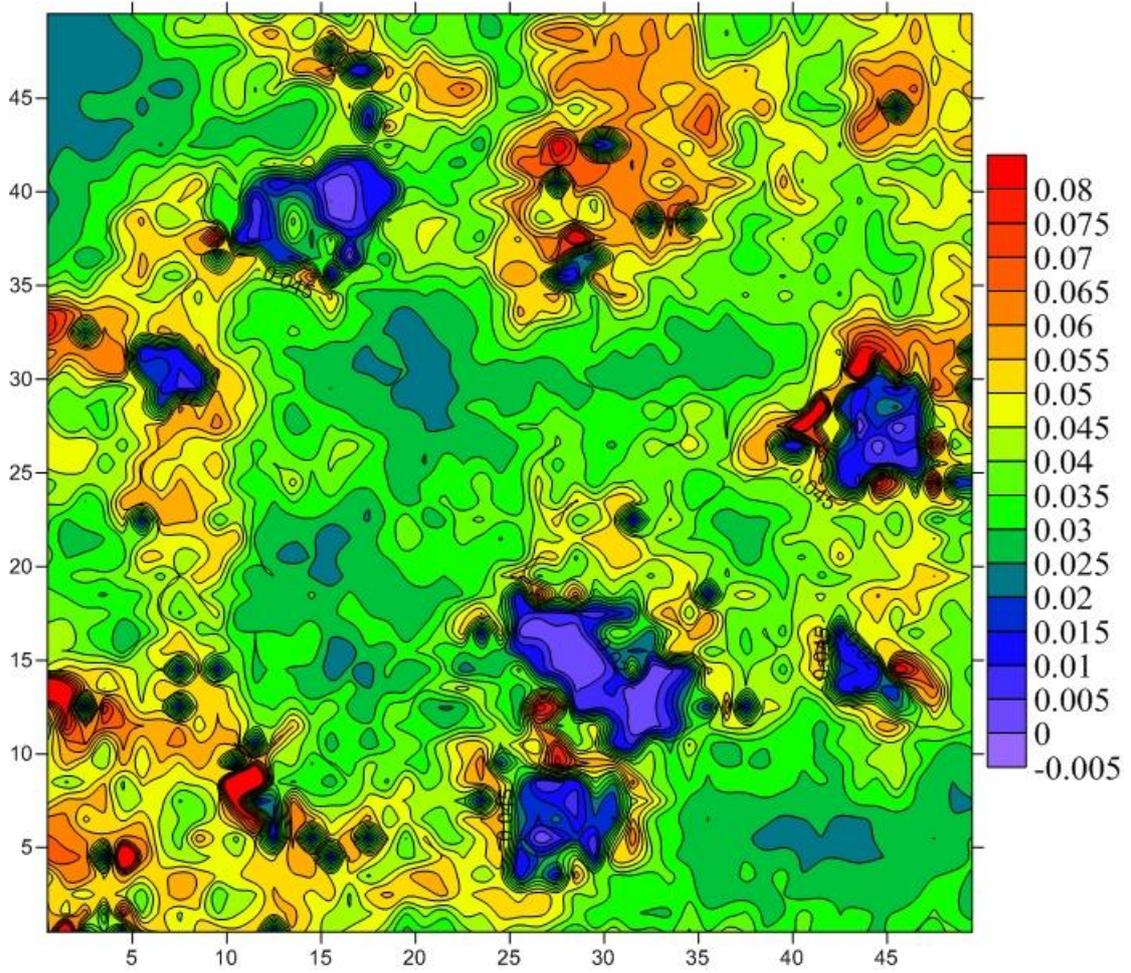

c. ΔH=1mm, t=4000h

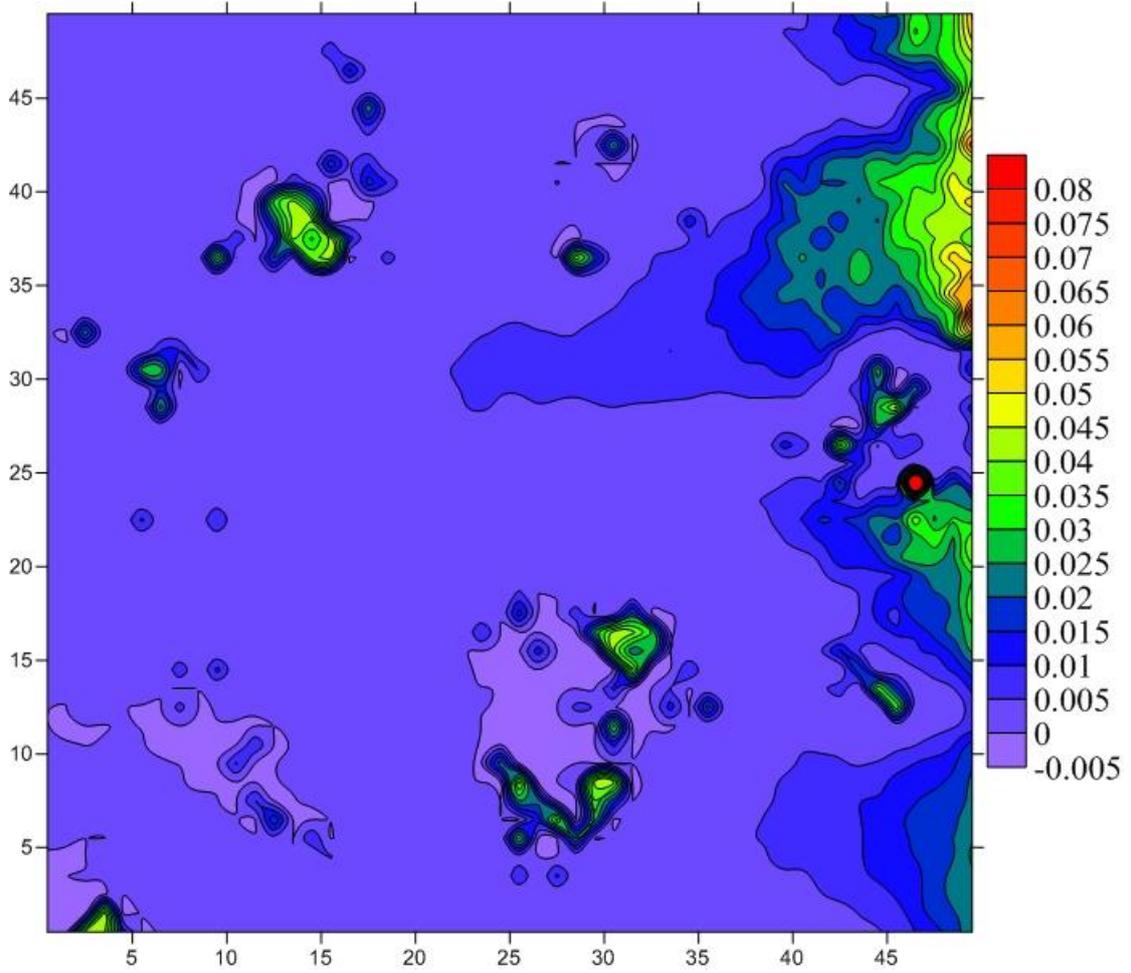

d. $\Delta H$=0.1mm, $t$=0

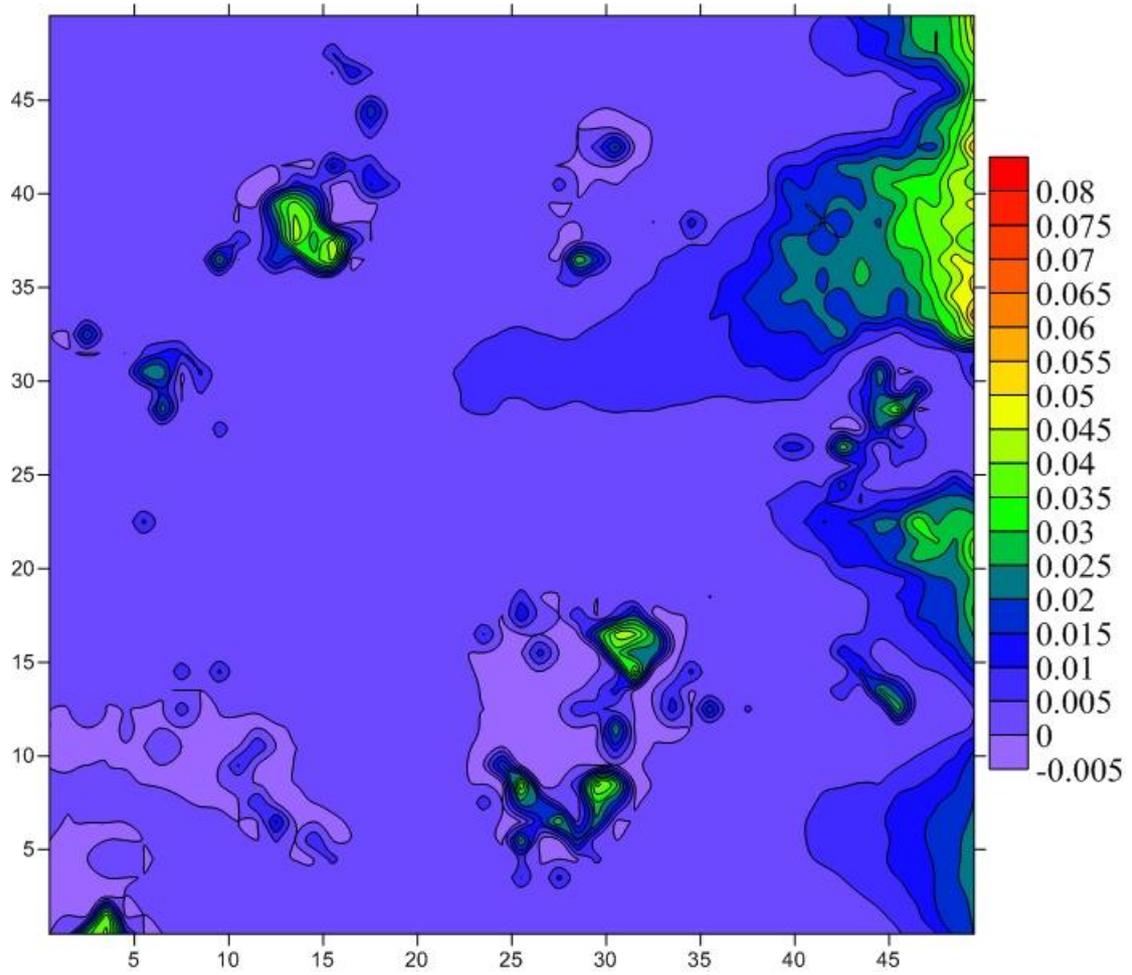

e. Δ*H*=0.1mm, *t*=2000h

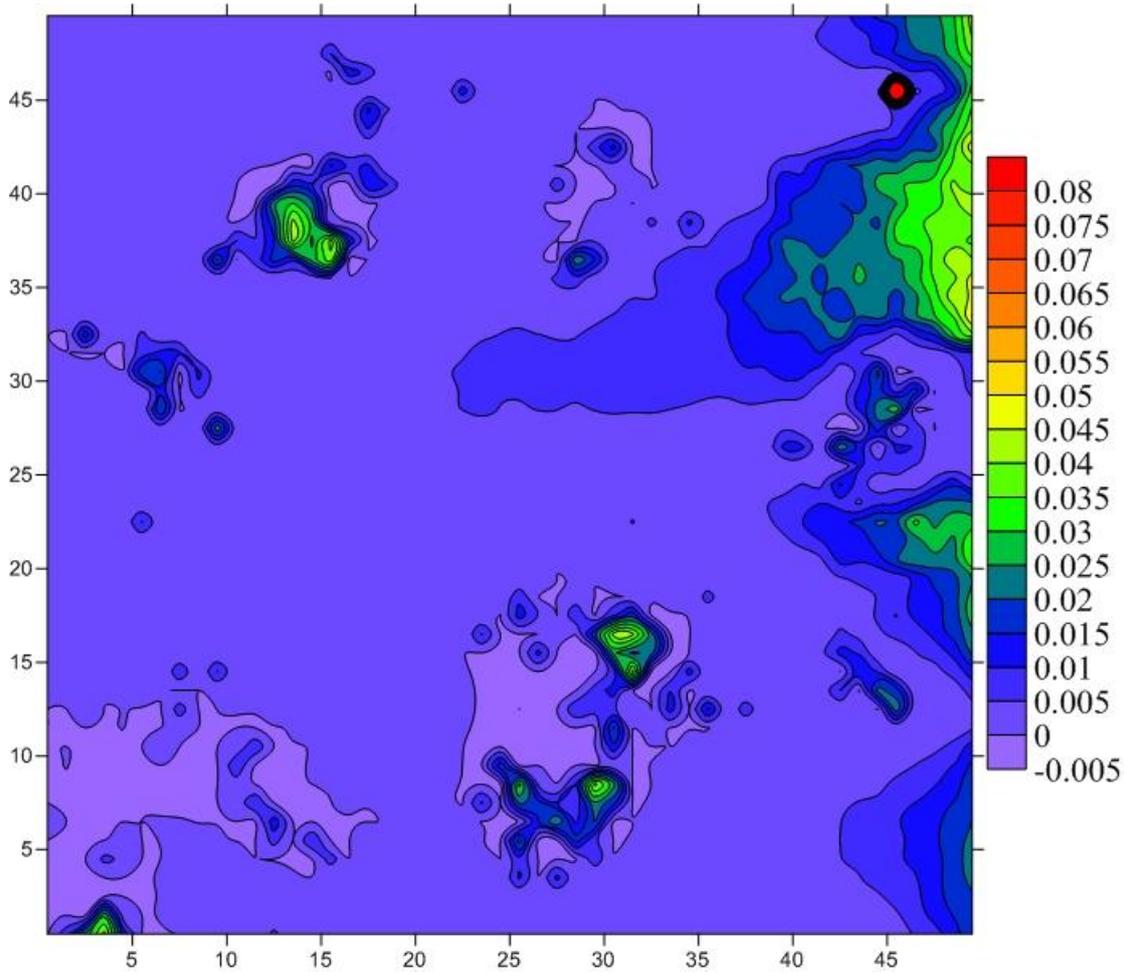

f. $\Delta H$=0.1mm, $t$=4000h

Figure 10. Dissolution rate alteration for flow fields with 0.1mm and 1mm hydraulic drawdowns

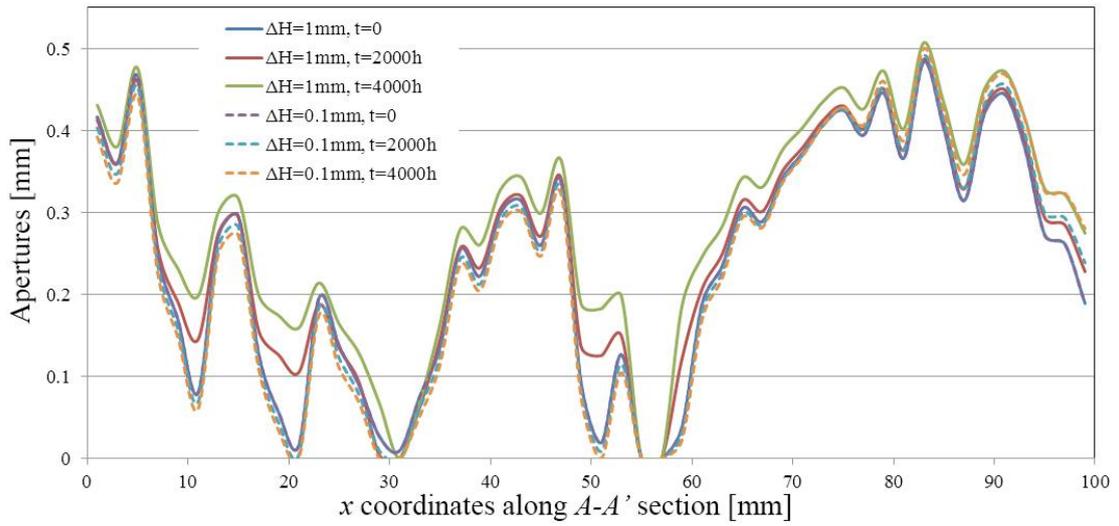

a. along *A-A'* section

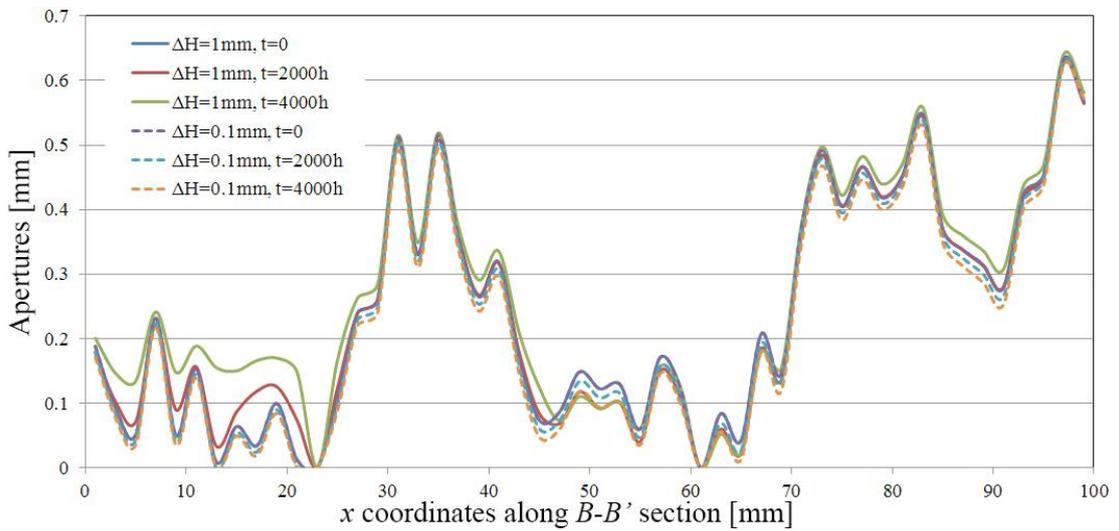

b. along *B-B'* section

Figure 11. Aperture alterations when hydraulic drawdown equals 0.1mm and 1mm.

(precipitation area expands continuously in dash lines)

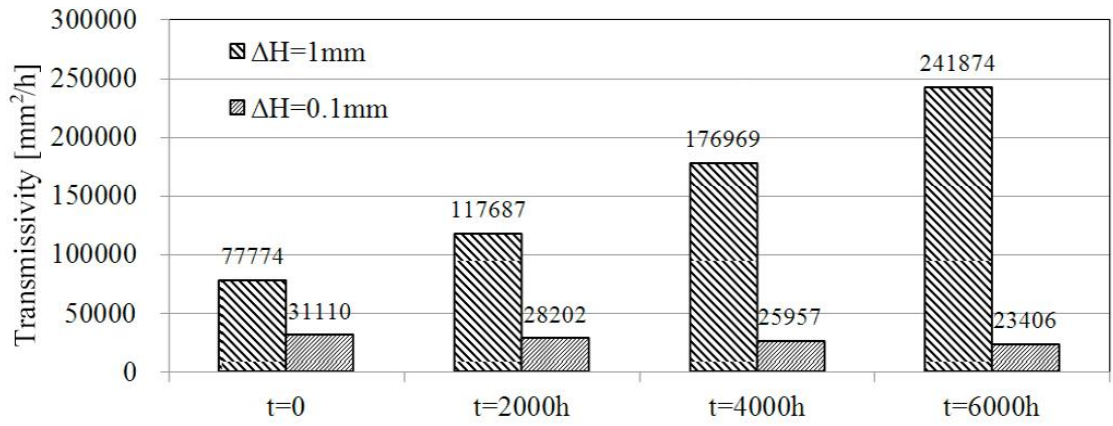

Figure 12. Fracture transmissivity alteration under hydraulic drawdown 1mm and 0.1mm.